\documentclass[a4paper,showkeys,floatfix,aps,pre,reprint,groupedaddress]{revtex4-1}
\usepackage{graphicx}
\usepackage{amsmath, amssymb, amsfonts}
\usepackage{color}
\usepackage{subfig}
\usepackage{caption}

\begin{document}

\title{Nucleation versus percolation: Scaling criterion for failure in disordered solids}



\author{Soumyajyoti Biswas}
\email[]{sbiswas@imsc.res.in (present address: ERI, Univ. of Tokyo)}
\author{Subhadeep Roy}
\email[]{sroy@imsc.res.in}
\author{Purusattam Ray}
\email[]{ray@imsc.res.in}
\affiliation{
The Institute of Mathematical Sciences, Taramani, Chennai-600113, India.\\
}

\date{\today}

\begin{abstract}
\noindent  One of the major factors governing the mode of failure in disordered solids 
is the effective range $R$, over which the stress field is modified following a
local rupture event. In random fiber bundle model, considered as a prototype of disordered 
solids, we show that 
the failure mode
is nucleation dominated in the large system size limit, as long as $R$ 
scales slower than $L^{\zeta}$, with $\zeta=2/3$. 
For a faster increase in $R$, the failure properties are dominated 
by the mean-field critical point, where the damages are uncorrelated in space. In that
limit, the precursory avalanches of all sizes are obtained even in the large system size limit. 
 We expect these results 
to be valid for systems with finite (normalizable) disorder.
\end{abstract}


\maketitle

Fracture is a complex phenomenon involving  large span of length and time scales.
From the ruptures at the micro or atomic level, fractures at the laboratory or engineering scales to 
earthquakes at the geological scales, the problem has attracted attention of the scientists 
for decades, cutting across disciplines \cite{books, phyrep, rmp}. 
A critical question in this problem is the abruptness of the failure process. A solid can fail following precursory rupture events or catastrophically without showing such precursors \cite{zapperi}.  The physical criterion that governs the mode of failure is an important open question. The  disorder present in the solid and the range of effective interaction (the range over which the stress field within the solid gets modified following a local rupture event) are the two major  factors that determine the mode of failure. 

The effect of disorder on the failure mode has been discussed recently.
Using random fuse model as a simple prototype of the disordered solid, it has been shown \cite{shk} that for finite disorder, the failure mode of the system, in the large system size limit, is always nucleation driven and therefore abrupt. The stress is nucleated around the largest defect and the defect grows in size until the the system fails. The precursory events (scale free size distribution of rupture events prior to failure etc.), previously seen in the model \cite{phyrep}, were attributed to 
the transient effect, implying these would not lead to the final fracture in the large system size limit. The only exception is the limit of extreme disorder \cite{herr}. However, experimentally such precursory features are observed (see e.g., \cite{corr}) for  which  the extreme disorder is not necessarily the physical condition. 

In this Communication, we ask the question: how does the range of the stress relief zone affect the mode of fracture 
in disordered solids? This range is an intrinsic property of the solid that depends on its elastic constants. 
We study fiber bundle model as a prototype of disordered solids and 
show that if the range $R$, over which the stress is released in unit 
time following a local rupture, is sufficiently large so that within the typical relaxation time (time taken for the stress values to come to a stable state following a rupture event), stress release takes place effectively over the entire system, the system can show scale-invariant precursory behaviour that survives even in the large system size limit. In particular, we find that if $R$ scales slower than a 
cut-off scale $R_c \sim L^{\zeta}$, with $\zeta=2/3$, where 
$L$ is the system size, the failure mode is nucleation dominated. On the other hand, if $R$ scales 
faster than $R_c$, the failure properties are dominated by the mean field behaviour even in the 
$L \rightarrow \infty$ limit and with finite disorder. In this case, local 
fiber ruptures are uncorrelated in space and the final failure is preceded by precursory avalanches, 
of rupturing fibers, of all sizes. $\zeta$ can be identified with the inverse of correlation length exponent. We apply the criterion to fiber bundle model with power-law stress redistribution \cite{inter1} and find the value of the exponent in the power law that demarcates the two limits. This demarcation is also supported by simulation results. 

The fiber bundle model has been studied extensively as it is a simple model which can capture  
various features of the failure of disordered solids \cite{first, dani, coleman, rmp1}. The model consists of a set of discrete elements (or fibers) each having a random failure threshold (mimicking the disorder in a solid) and fixed between a rigid ceiling and a bottom plate. On application of a stress (loading the plate), some of the elements may fail, which raises the effective stress on the remaining elements, some of which may now fail and so on.  It is known that in one extreme limit of the model, where the failure of one element affects the stability of all others equally (equal load sharing model; when the bottom plate is absolutely rigid), the failure mode is precursor driven and the damage is diffused in the system in the sense that it occurs all over the system in an uncorrelated manner. In another limit, where the failure of one element affects only the element(s) nearest to it (local load sharing model; bottom plate is absolutely soft), the failure is nucleation driven \cite{harlow}. Both these situations are, however, far from reality. The first case implies the absence of any notion of distance in the system, hence excluding the concept of stress nucleation altogether. The second limit indicates a very low elastic modulus, which is also physically unrealistic.  
There have been previous attempts \cite{inter1,inter2,inter3,inter4} to interpolate between these two extreme limits of the model, but they did not arrive at any general criterion for which the crossover is observed.

We take here a linear array of fibers and set a redistribution rule for the load of a failed fiber such that the range over which the stress is redistributed has a scale $R$. 
We follow the rule, whereby, the load on a broken fiber is distributed uniformly among the $R$ 
successive surviving fibers (see also \cite{inter2r}) on either side of the broken fiber. 
The system is loaded gradually until it fails completely.
The disorder is modelled by random failure thresholds of the fibers, which are drawn from a uniform distribution in $[0:1]$.  
 In one scan of the lattice (one time step), all fibers having load more than their thresholds are 
 broken. The load of these fibers are redistributed according to the redistribution rule mentioned 
 above. 
 If the load on any one of these neighbors exceeds threshold after redistribution, it is broken in the next scan. This continues until
no fiber is broken in a given scan. The external load is then increased just upto the point when the weakest of the remaining surviving fiber breaks
and the above dynamics is continued.
 
\begin{figure}[t]
\centering 
\captionsetup{justification=raggedright}
\includegraphics[width=9cm]{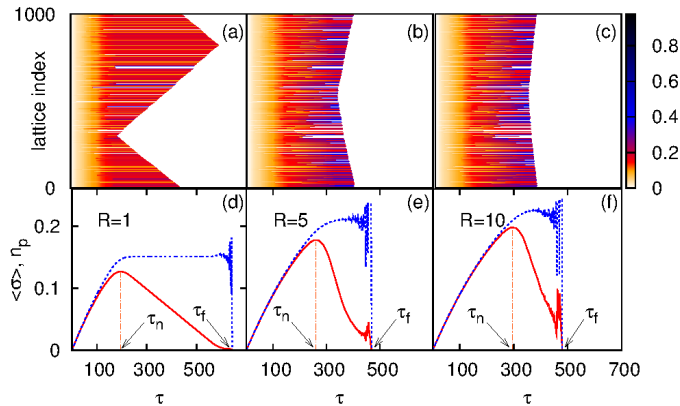}
   \caption{(Color online) The upper panel shows the variation of the stresses on the fibers in the system with redistribution steps $\tau$. 
The onset of nucleation can be seen from the tip of the cone, beyond which one broken patch grows
and the total number of patch starts decreasing. The onset time gets shifted to a higher value as the stress redistribution range $R$ is increased ((a)-(c))
and expected to merge with failure time in the global load sharing limit. The panel below ((d)-(f)) shows the variation of
average stress per fiber ($\langle \sigma\rangle$; denoted by dotted lines) and number ($n_p$) of patches (denoted by solid lines), scaled by system size, 
for corresponding $R$ values. 
The onset of nucleation ($\tau_n$) is the point where the
number of patch starts decreasing and the stress per fiber saturates (at the critical value) until failure point ($\tau_f$) is reached.}
\label{config}
\end{figure}

Consider the case when $R$ is constant, i.e. it does not scale with the system size $L$. 
This is a generalization of the usually studied local load sharing model ($R=1$). For threshold distributions that
extend to zero (e.g., uniform distribution in [$0:1$]), for an arbitrarily small applied load $\sigma$, there will be a large
enough patch of length $m$ of successive broken fibers, such that the redistributed stress on its 
surviving neighbours ($R$ on either sides) will exceed the maximum threshold value ($1$ in this case) \cite{rmp1} leading to the catastrophic failure of the system. It was shown that for $R=1$, $\sigma_c\sim 1/\ln L$ \cite{rmp1} where $\sigma_c$ is the critical value of stress for which the system fails.
For arbitrary $R$, one would expect $m\sim R$. But as long as $R$ does not scale with $L$, the qualitative behaviour is found (numerically) to remain the same.

\noindent A more interesting case is when the $R$ is varied to see the effect on the failure mode. Obviously, for 
$R\sim L$, the load redistribution becomes global by definition and the mode of failure is expected to be gradual with
usual avalanche statistics. The question we intend to answer is: Does the scaling of the effective range have to be 
as fast as linear to lead to global load sharing failure mode?

A signature of the global load sharing process is the uncorrelated failure of fibers when load is increased. This leads to 
creation of new broken patches in the system with the increase of load. On the other hand, onset of nucleation is essentially 
the growth of one patch
that engulfs all other patches, leading to the failure of the system. An effective way to detect nucleation, therefore, is 
to monitor the number of broken patches in the system. The top panel of Fig. \ref{config} shows the evolution of the load 
per fiber with time (defined here as the number of load redistribution step). 
As can be seen from Fig. \ref{config} (a)-(c), for different values of $R$, upto the onset of nucleation (tip of the cone), 
each fiber carries almost the same load. When nucleation sets in, one single patch starts growing  
leading to the complete failure of the system. 
As can be seen from the bottom panel, the time $\tau_n$ of onset of nucleation is where the 
number of patches (scaled by system size) starts decreasing. It is
also the time when the load per fiber value becomes constant (implying that to be the critical load). After many steps of
load redistribution (each redistribution considered here as one time step), the system finally fails completely at time $\tau_f$.

\begin{figure}[t]
\centering
\captionsetup{justification=raggedright}
 \includegraphics[width=9cm]{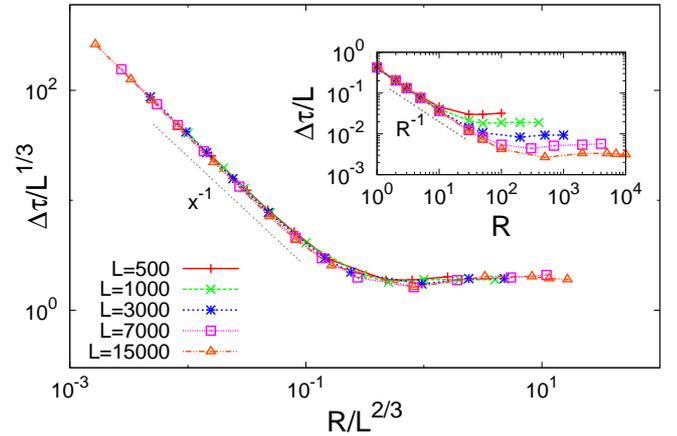}
   \caption{(Color online)The data collapse for the time difference $\Delta\tau=\tau_f-\tau_n$ for different 
system sizes $L$ and for different range $R$ as given by the 
scaling form Eq. (\ref{scaling_form}). The inverse decay marks the nucleation regime, which stops
when $R\sim L^{2/3}$ and global load sharing region begins. In this regime, $\Delta \tau$ becomes $R$ independent but depends on $L$ (as can be
seen from the inset). The initial collapsed region in the inset confirms the dependence $\Delta \tau\sim L/R$
in the nucleation regime, and the lines spreads out as soon as global mode starts dominating.}
\label{scaling_time}
\end{figure}

As the range of load sharing is increased, the nucleation and failure times approach each other i.e., $\Delta \tau=\tau_f-\tau_n$
decreases. While the order of these events can not be reversed, they may come very close (upto a scale of critical relaxation time in mean field limit) as $R$ increases, 
implying vanishing of the nucleation mechanism. In Fig. \ref{scaling_time} (inset) the variation of the time interval $\Delta \tau$ is shown
with $R$ for different system sizes.  It shows a initial linear decrease, followed by a saturation regime, which can be interpreted
as the vanishing of nucleation mechanism. The value of the saturation time depends on the system size. We find an overall scaling form 
\begin{equation}
\Delta \tau\sim L^{\alpha}\mathcal{F}\left(\frac{R}{L^{\zeta}}\right).
\label{scaling_form}
\end{equation}
Satisfactory data collapse is obtained for $\alpha=0.33\pm 0.01$ and $\zeta=0.66\pm 0.01$ (see Fig. \ref{scaling_time}), which leads to the conjectured 
exact values as $\alpha=1/3$ and $\zeta=2/3$. The scaling function $\mathcal{F}(x)$ has the form $\mathcal{F}(x)\sim 1/x$ for $x<1$ and
$\mathcal{F}(x)$ becomes constant for $x\ge 1$. 

Before interpretation of the consequences of such scaling form, let us try to
understand the exponent values. For small values of $R$, the nucleation sets in from the weakest patch, where the rest of the system is almost
intact. The patch then grows, breaking $2R$ neighbours on each step of redistribution, until the whole system breaks. The time required for
complete failure should be $\Delta \tau\sim L/v_f$, where $v_f$ is the growth velocity of the fatal patch and it has to cover almost the 
entire lattice, hence the numerator $L$. Now, as $\sim2R$ fibers break in each step, $v_f\propto R$, giving $\Delta \tau\sim L/R$. 
This is what is seen in the early part of the scaling. Now, the part where $\Delta\tau$ is independent of $R$, the failure mode is of 
global load sharing type, where the relaxation time at the failure point diverges as $\Delta \tau\sim L^{\alpha}$, with $\alpha=1/3$ \cite{explain, manna}. 
Therefore, for matching of the two scaling forms at the crossover one must have $L^{\alpha}(R/L^{\zeta})^{-1}\sim L/R$, giving
$\alpha+\zeta=1$. Therefore, $\zeta=2/3$ as is also seen from data collapse.  

To understand the physical picture, let us consider the probability distribution of stress values within the system. For a global load sharing 
model each fiber carry same stress, hence the distribution function is a delta function.
On the other hand, for stress nucleation (and failure driven by it), the stress distribution function must have a finite width that survives 
the large system size limit. The width depends on the (i) the range $R$ of the stress release in unit time (the width is narrower as $R$ becomes larger)
and (ii) the fluctuation in the number of broken fibers, which contributes in the increase of the width. The functional dependence 
of the width is expected to be of the form $\Delta \sigma\sim \Delta N/R$, where $\Delta N$ is the fluctuation in the number of surviving
fibers. Since we are approaching the mean field critical point, the relevant fluctuation is the one seen near it. But
it is known that the fluctuation in the fraction of surviving fibers (over the disorder configurations)
 scales as $\Delta p_c\sim L^{-1/3}$ \cite{inter3, daniels2}, thus
the fluctuation in the number will scale as $\Delta N\sim L^{2/3}$. Therefore $\Delta \sigma \sim L^{2/3}/R$. Hence $\Delta \sigma$ retains a 
finite value in the large system size limit only when $R=R_c\sim L^{2/3}$ (which is seen from the scaling relation Eq. (\ref{scaling_form})). For $R > R_c$, the stress distribution is narrow, which corroborates to 
the absence of stress nucleation. This sets the scaling criterion for nucleation.

\begin{figure}[ht]
\begin{center}

\includegraphics[scale=0.5]{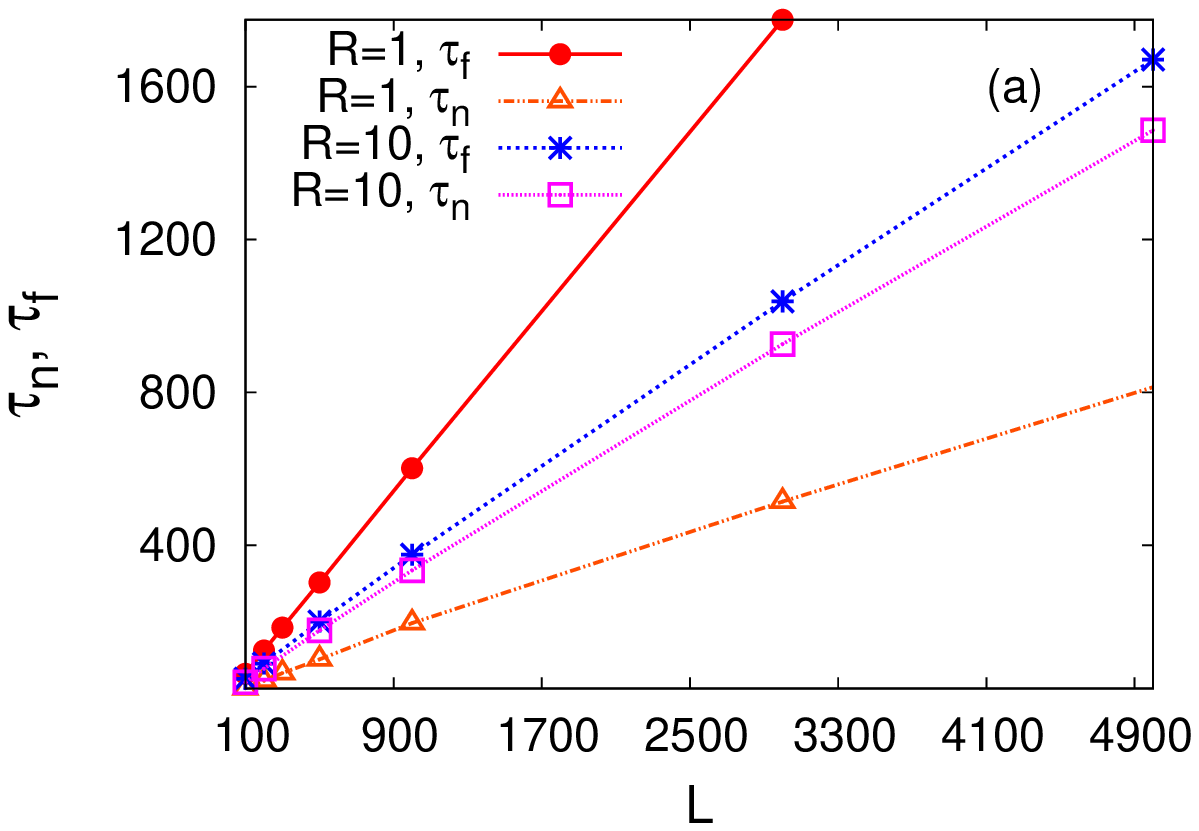}
\includegraphics[scale=0.5]{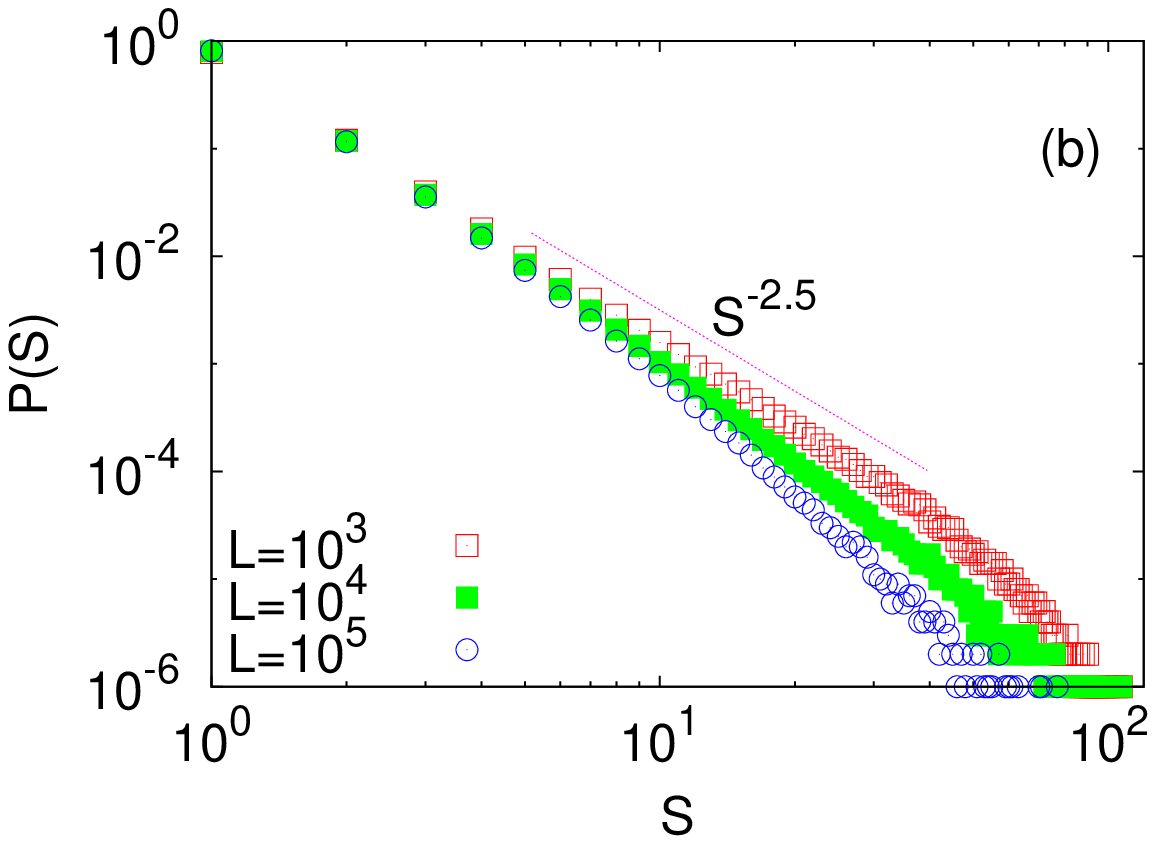}
\end{center}
\captionsetup{justification=raggedright}
\caption{(Color online) The top figure (a) shows the variation of the onset time for nucleation $\tau_n$ and failure time $\tau_f$ with system size $L$ for two values of $R$. 
For small system sizes and bigger $R$ value, the two times are very close to each other. In this region,  
mean-field `critical
behaviour' can be observed. For fixed $R$ value, this `criticality' will not survive in the large system size limit. The figure at the
bottom (b) shows the avalanche size distribution for fixed $R$ value ($100$) while the system size is increased. For $L=10^3$, the ratio
$R/L^{2/3}=1$, putting the system in the critical regime where $P(S)\sim S^{-2.5}$. But for $L=10^4$ and $L=10^5$ the ratio
becomes $\approx 0.215$ and $\approx 0.046$ respectively. The avalanche size distribution
in the last two cases deviates from the above scale free distribution.} 
\label{scaling_diff}
\end{figure}

 The phrase `large system size limit' is very important in the above statement, 
since for a given choice of $(R,L)$ the system may show scale free avalanche distribution, which may
go away for large system size $L > R^{3/2}$ when the fracture mode becomes 
nucleation dominated. As can be seen from Fig. \ref{scaling_diff} (a), when the system size is small,
for a given $R$, $\tau_n$ and $\tau_f$ are very close. In that region, the avalanche size distribution (Fig. \ref{scaling_diff} (b)) and other related
quantities shows mean field behaviour. But as the system size is increased, the critical behaviour goes away. This is similar to
the `finite size criticality' mentioned in Ref. \cite{shk}. However, in our case we can tune the range of stress release and 
the mean-field like critical behaviour survives in the thermodynamic limit provided the range $R$ increases sufficiently fast (although sub linearly) 
with system size. 

\begin{figure}[t]
\centering 
\captionsetup{justification=raggedright}
\includegraphics[width=8cm]{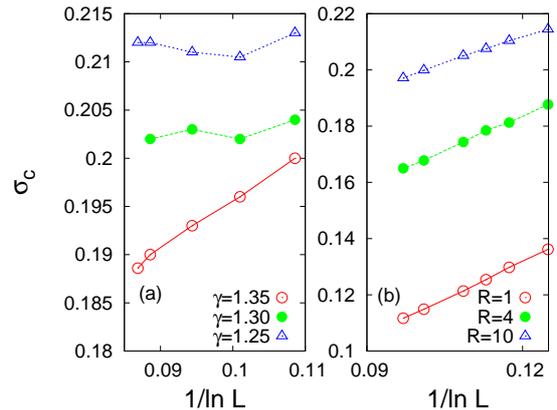}
   \caption{(Color online) The left figure (a) shows that the failure threshold $\sigma_c$ goes to zero with $1/\ln L$ when $\gamma<\gamma_c=4/3$ in the case
of power-law load redistribution, showing the validity of the proposed scaling criterion that the effective range must scale 
faster than $L^{2/3}$ (or $\gamma<4/3$) for global load sharing mode. The inverse logarithmic decay is similar to what is
observed for uniform load sharing within range $R$ for $R<L^{2/3}< 1$ (b).}
\label{scaling_log}
\end{figure}
We expect that the scaling is valid for other forms of load redistribution where one has a
characteristic length scale. We have checked this for several cases such as exponential decay, linear decay etc.
An interesting question is what happens for a `scale free' redistribution rule, e.g  power-law redistribution? In this case, the load redistributed on the
$j$-th fiber after the failure of $i$-th fiber is proportional to $1/|i-j|^{\gamma}$ \cite{inter1}. The distribution is not truly scale free, because
it has two cut-offs, viz the lower cut-off due to lattice spacing (which we take as unity) and a upper cut-off due to finite system size.
A quantity `average range of interaction' will interpolate between these two extremities as $\gamma$ is tuned. Remembering the normalization
$A\int\limits_1^Ldx/x^{\gamma}=$ constant, one can always calculate the average (effective) length of interaction as 
$R^{eff}=\langle x\rangle\sim \frac{1-\gamma}{2-\gamma}\frac{L^{2-\gamma}-1}{L^{1-\gamma}-1}$. Clearly for $\gamma<1$, $R^{eff}\sim L$ for large
$L$. Also, $R^{eff}\to $ constant for $\gamma>2$ implying nucleation scenario with fixed ranged interaction. But for $1<\gamma<2$, $R^{eff}\sim L^{2-\gamma}$ for $L\to\infty$. According to the 
scaling argument presented
above, the critical value of the exponent $\gamma_c$ for which the failure behaviour crosses over from nucleation to global 
load sharing mode is to be given by $2-\gamma_c=2/3$ or $\gamma_c=4/3$. In support of this claim, in Fig. \ref{scaling_log} we have shown 
the behaviour of critical load $\sigma_c$ for fracture for various $\gamma$ values. It is seen that for $\gamma>\gamma_c=4/3$, $\sigma_c\sim 1/\ln L$, which
is similar to what we see in the nucleation regime for uniform load redistribution among $R$ neighbours with $R/L^{2/3}\to 0$ as $L\to\infty$. On the other hand, 
$\sigma_c$ saturates to a non-zero value when $\gamma<4/3$, as is expected in global load sharing scheme.   

In conclusion, we have obtained a scaling criterion for the effective range of stress release in unit time  for which
the fracture mode of a  disordered system crosses over from nucleation dominated regime to percolation like (uncorrelated) failure with mean-field critical
behaviour.   We have studied random fiber bundle model as a simple prototype of a disordered system. The failure phenomena in less simplified models such as random fuse or random spring 
networks are expected to converge to that of fiber bundle models in higher dimensions \cite{sinha}. Therefore, specifying a criterion of the crossover in this
simplest case can help in formulating the same for more realistic systems.


\begin{thebibliography}{99}

\bibitem{books}
{\it Statistical Models for the Fracture of Disordered Media}, Eds H. Herrmann and S. Roux 
(North-Holland, Amsterdam, 1990); B. K. Chakrabarti and L. G. Benguigui, {\it Statistical Phtsics
of Fracture and Breakdown in Disordered Systems} (Oxford Univ. Press, Oxford, 1997); M. Sahimi,
{\it Heterogeneous Materials II: Nonlinear and Breakdown Properties} (Springer-Verlag, New York, 2003).
\bibitem{phyrep}
M. J. Alava, P. K. V.V. Nukala, S. Zapperi, Adv. Phys. {\bf 55}, 349 (2006).
\bibitem{rmp}
H. Kawamura, T. Hatano, N. Kato, S. Biswas, B. K. Chakrabarti, Rev. Mod. Phys. {\bf 84}, 839 (2012).
\bibitem{zapperi}
S. Zapperi, P. Ray, H. E. Stanley, and A. Vespignani, Phys. Rev. Lett. {\bf 78}, 1408 (1997).
\bibitem{shk}
A. Shekhawat, S. Zapperi, J. P. Sethna, Phys. Rev. Lett. {\bf 110}, 185505 (2013).

\bibitem{herr}
A. A. Moreira, C. L. N. Oliveira, A. Hansen, N. A. M. Ara\'{u}jo, H. J. Herrmann, J. S. Andrade, Jr. Phys. Rev. Lett. {\bf 109}, 255701 (2012).


\bibitem{corr}
J. Bar\'{o}, \'{A}. Corral, X. Illa, A. Planes, E. K. H. Salje, W. Schranz, D. E. Soto-Parra, E. Vives, Phys. Rev. Lett. {\bf 110}, 088702 (2013).
\bibitem{first}
F. T. Peirce, J. Text. Inst. {\bf 17}, T355 (1926).

\bibitem{dani}
H. E. Daniels, Proc. R. Soc. London, Ser. A {\bf 183}, 405 (1945).

\bibitem{coleman}
B. D. Coleman, J. Appl. Phys. {\bf 27}, 862 (1956).

\bibitem{rmp1}
 S. Pradhan, A. Hansen, B. K. Chakrabarti, Rev. Mod. Phys {\bf 82}, 499 (2010).
\bibitem{harlow}
D. G. Harlow, S. L. Phoenix, J. Composite Mater. {\bf 12}, 195 (1978);
D. G. Harlow, S. L. Phoenix, J. Mech. Phys. Solids, {\bf 39}, 173 (1991).

\bibitem{inter1}
R. C. Hidalgo, Y. Moreno, F. Kun, H. J. Herrmann, Phys. Rev. E {\bf 65}, 046148 (2002).

\bibitem{inter2}
S. Pradhan, B. K. Chakrabarti, A. Hansen, Phys. Rev. E {\bf 71}, 036149 (2005).
\bibitem{inter3}
A. Stormo, K. S. Gjerden, A. Hansen, Phys. Rev. E {\bf 86}, 025101(R) (2012).

\bibitem{inter4}
 S. Biswas, B. K. Chakrabarti, Eur. Phys. J. B {\bf 86}, 160 (2013).





\bibitem{inter2r}
F. Kun, S. Zapperi, H. J. Herrmann, Eur. Phys. J. B {\bf 17}, 269 (2000).

\bibitem{explain}
The fraction of surviving fibers (in global load sharing model) at $\sigma_c$ decays with time as $U(t)=\frac{1}{2}\left(1+\frac{1}{t+1}\right)$ \cite{rmp1}.
Hence after a time $\Delta\tau$ (relaxation time), the deviation from the critical surviving fraction ($1/2$) will be $1/2(\Delta\tau+1)$. But this deviation
scales as \cite{inter3, daniels2} $\Delta p_c\sim L^{-1/3}$, giving $1/2(1+\Delta\tau)\sim L^{-1/3}$. In the large $\Delta\tau$ limit, $\Delta\tau\sim L^{1/3}$, which is 
what is also seen numerically in \cite{manna}.



\bibitem{manna}
C. Roy, S. Kundu, S. S. Manna, Phys. Rev. E {\bf 87}, 062137 (2013).

\bibitem{daniels2}
H. E. Daniels and T. H. R. Skyrme, Adv. Appl. Probab. {\bf 21}, 315 (1989).

\bibitem{sinha}
S. Sinha, J. T. Kjellstadli, A. Hansen, arXiv:1501.02489 (2015).

\end{thebibliography}
\end{document}